\documentclass[preprint,review,12pt]{elsarticle}

\usepackage[margin=3cm]{geometry}

\usepackage[noend]{algpseudocode}
\usepackage[english]{babel}
\usepackage{verbatim}
\usepackage[table,xcdraw]{xcolor}
\usepackage{graphicx}
\usepackage{subcaption} 
\usepackage{amssymb}
\usepackage{amsmath}
\usepackage{amsfonts}
\usepackage{mathrsfs}
\usepackage{mathtools}
\usepackage{float}
\usepackage{multirow}
\usepackage{tabularx}
\usepackage{array}
\usepackage{color}
\usepackage{leftidx}
\biboptions{numbers,sort&compress}
\usepackage{xcolor}
\usepackage[]{algorithm2e}
\usepackage{textgreek}
\usepackage[colorlinks,citecolor=red,urlcolor=blue,bookmarks=false,hypertexnames=true]{hyperref} 
\usepackage{comment}

\usepackage{booktabs}
\definecolor{MC}{rgb}{0.0, 0.5, 0.0}

\newcommand*{\colorboxed}{}
\def\colorboxed#1#{%
  \colorboxedAux{#1}%
}
\newcommand*{\colorboxedAux}[3]{%
  \begingroup
    \colorlet{cb@saved}{.}%
    \color#1{#2}%
    \boxed{%
      \color{cb@saved}%
      #3%
    }%
  \endgroup
}

\usepackage{dashbox}

\usepackage{tikz}
\usepackage{soul}

\usepackage{rotating}

\usepackage{lineno}

\newcommand{\RN}[1]{%
  \textup{\uppercase\expandafter{\romannumeral#1}}%
}

\definecolor{azure(colorwheel)}{rgb}{0.0, 0.5, 1.0}

\definecolor{mygreen}{rgb}{0.0, 0.5, 0.0}
\definecolor{darkraspberry}{rgb}{0.53, 0.15, 0.34}
\definecolor{bleudefrance}{rgb}{0.19, 0.55, 0.91}

\newcolumntype{C}[1]{>{\centering\arraybackslash}m{#1}}

\newcommand{\eg}{\textit{e}.\textit{g}., }
\newcommand{\ie}{\textit{i}.\textit{e}. }

\begin{document}

\begin{frontmatter}


\title{Test-time data augmentation: improving predictions of recurrent neural network models of composites}

\author[]{Petter Uvdal}
\address[]{Department of Physics, University of Gothenburg, Origovägen 6B, 41296 Gothenburg, Sweden}

\author[]{Mohsen Mirkhalaf\corref{mycorrespondingauthor}}
\cortext[mycorrespondingauthor]{Corresponding author}
\ead{mohsen.mirkhalaf@physics.gu.se}
\ead[url]{www.materialslab.org}

\begin{abstract}
Recurrent Neural Networks (RNNs) have emerged as an interesting alternative to conventional material modeling approaches, particularly for non-linear path-dependent materials. Remarkable computational enhancements are obtained using RNNs compared to classical approaches such as the computational homogenization method. However, RNN predictive errors accumulate, leading to issues when predicting temporal dependencies in time series data. This study aims to address and mitigate inaccuracies induced by neural networks in predicting path-dependent plastic deformations of short fiber reinforced composite materials. We propose using an approach of Test-Time data Augmentation (TTA), which, to the best of the authors' knowledge, is previously untested in the context of RNNs. The method is based on augmenting the input test data using random rotations and subsequently rotating back the predicted output signal. By aggregating the back-rotated predictions, a more accurate prediction compared to individual predictions is obtained. Our analysis also demonstrates improved shape consistency between the prediction and the target pseudo time-signal. Additionally, this method provides an uncertainty estimation which correlates with the absolute prediction error. The TTA approach is reproducible with different randomly generated data augmentations, establishing a promising framework for optimizing predictions of deep learning models. We believe there are broader implications of the proposed method for various fields reliant on accurate predictive data-driven modeling.
\end{abstract}


\end{frontmatter}
\section{Introduction}
\label{Introduction}
Traditionally, low-fidelity mean-field models (\eg Eshelby \cite{Eshelby1957}, Hashin and Shtrikman \cite{HASHIN1962335, HASHIN1963127}, Hill \cite{Hill1965}, Budiansky \cite{BUDIANSKY1965223}, and Mori-Tanaka \cite{Mori1973}) have been used to model short fiber reinforced composites (SFRCs). However, to more accurately capture the elasto-plastic behavior of SFRCs, more computationally expensive high-fidelity full-field models (\eg Finite Element Method (FEM) and Fast Fourier Transform (FFT)) \cite{Qi2015, Spahn2014, Schneider2016, Mirkhalaf2020} have been used. Yet, the following challenges remain: (i) difficulty in generating different Representative Volume Elements (RVEs) which mimic the actual material micro-structure \cite{Hoang2016, Harper2012}, and (ii) high computational expense \cite{Bargmann2018, Mirkhalaf2022}. Furthermore, due to the stochastic stacking of fibers in RVE realizations, different RVEs exist for the same orientation tensor and fiber volume fraction, resulting in a variation in the output of the simulations. These issues highlight the demand for more dependable and efficient models.

In recent years, Artificial Neural Networks (ANNs) have emerged as an alternative to classical numerical simulations with remarkably lower computational requirements \cite{Mozaffar2019, Wang2018, Ghane2023, Bonatti2022, Liu2023}. For modeling plasticity, RNN models have been implemented which showed a strong capability to  predict the non-linear path-dependent behavior \cite{Wang2018, Mozaffar2019, Wu2020}. In addition to RNNs, other neural network architectures have been developed that embed constitutive models within deep learning frameworks \cite{Maia2023} (for a review, see \cite{LIU2021}). However, classical RNN models face important challenges, such as data scarcity, extrapolating predictions outside of the region of the training dataset, and lack of reliable uncertainty estimates \cite{Bhadeshia2009}. To have an overview of recent developments, challenges, and potential future perspectives, see \cite{Mirkhalaf2024}.

In terms of uncertainty, several machine learning methods have been developed to obtain the uncertainty probability distribution for various applications, \ie Hamiltonian Monte Carlo (HMC) methods \cite{neal1996bayesian}, variational inference \cite{Hinton1993, NIPS2011_7eb3c8be}, deep ensembles \cite{Balaji2017}, dropout \cite{Srivastava2014}, Stochastic Weight Averaging (SWA) \cite{izmailov2019}, and randomized prior networks \cite{NEURIPS2018_5a7b238b, YANG2022115399}. However, these methods are not without limitations, \eg deep ensembles (for example bootstrapping) requires training of multiple networks and architectures and do not take into account prior information. Variational inference, which randomly drops out neurons in the neural network, can produce inaccurate uncertainty estimates in simple neural networks \cite{pearce2018}. Bayesian neural networks (BNNs) make use of HMC methods. In BNNs, uncertainty is classified into two types: \textit{epistemic} uncertainty, resulting from ignorance due to data scarcity, and \textit{aleatoric} uncertainty, which contains inherent noise in the data \cite{OLIVIER2021114079}. HMC method can accurately calculate the posterior probability distribution while taking into account the prior information. Yet, HMC suffer from high computational cost and typically are difficult to scale in relation to the number of parameters and data points \cite{moya2022, papamarkou2021}. Since uncertainty estimates are crucial for practical applications, more methods of quantifying uncertainty are required in the field.

Related to uncertainty, is the accuracy of the ANNs. Their data-hungry nature in combination with inefficient physics-based simulations results in a data scarcity issue, leading to inaccurate predictions. Different approaches have been introduced to address data scarcity: (i) efficient micro-mechanical approaches \cite{Mentges2021, RAO2020}, (ii) transfer learning \cite{Cheung2024_transfer, GHANE2024105378, XU2023} (also see meta-learning as a related method \cite{finn2017, CHENG2022, JEONG2024}), and (iii) training-time data augmentation \cite{Cheung2024, DONG2022117541, Sajjadinia2024, Kim2021_design, Shorten2019, MUMUNI2022100258}, (iv) physically or mechanistically-informed neural networks \cite{Liu2023}. In addition to data scarcity issue, another important challenge in ANN model developments is the network hyperparameter optimization which is typically done via either a trial-and-error or a grid-search procedure. A recent study \cite{nikzad2024noveltaguchibasedapproachoptimizing} proposed an efficient approach using the Taguchi design of experiments method for hyperparameter optimization. While these methods are promising, in the field of SFRCs, data augmentation has recently been implemented as a novel method to improve accuracy \cite{Cheung2024}.

Training-time data augmentation is an established technique in various fields, such as in image analysis \cite{Shorten2019, MUMUNI2022100258} and language processing \cite{Shorten2021}. For modeling of SFRCs, training-time data augmentation has recently been applied to augment a training dataset of full-field surrogate models, by augmenting tensors using rotations \cite{Cheung2024}. Other related works in the field of computational mechanics involve performing 3D transformations on Computer-Aided Design (CAD) parts \cite{DONG2022117541}, and by interpolating in spatio-temporal properties in multiscale FE-models \cite{Sajjadinia2024}. While training-time data augmentation effectively addresses data scarcity by enriching training datasets, individual predictions could remain inaccurate. In the previous RNN models developed for SFRCs, there are particularly inaccurate predictions when the output stress is near zero \cite{Cheung2024, Cheung2024_transfer}. This noise might be due in part to incomplete training resulting from scarce data or an inherent property of the neural network architecture, which, to the best of the authors' knowledge, remains unresolved.

In this paper, we propose the following approach: to implement test-time data augmentation (TTA) in an RNN to (i) provide more accurate predictions with less inherent noise and (ii) obtain an uncertainty measure of the prediction relative to the target. As opposed to training-time data augmentation, TTA focuses on enhancing prediction accuracy during inference by reducing prediction artifacts without modifying the ANN model or its training data \cite{Moshkov2020}. TTA, as the name suggests, refers to augmenting a test-time dataset, to produce multiple ANN predictions from a single input. Thus, by analyzing the multiple predictions, a more robust prediction can be obtained. Consequently, TTA provides an approach capable of quantifying prediction variation and reducing individual prediction artifacts. Previously, TTA has been used to enhance deep learning models in various applications, \eg segmenting microscopy images, determining aleatoric uncertainty in medical image segmentation, and improving the robustness of deep-learning models used for underwater acoustic signal classification \cite{Moshkov2020, Xu2024, Wang2019}. In this study, we are following the method developed by Cheung et al. \cite{Cheung2024}, who proposed training-time data augmentation for SFRCs, to develop a TTA method.

The structure of the rest of this paper is as follows. Section \hyperlink{section.2}{2} presents a summary of our previous relevant studies on the generation of different datasets, RNN model developments, and techniques developed to address data scarcity issues. Section \hyperlink{section.3}{3} explains the data augmentation approach during inference. Section \hyperlink{section.4}{4} discusses results and implications, covering the variance of prediction using the TTA method, and its effect on prediction error, shape consistency, and uncertainty estimates. Section \hyperlink{section.5}{5} concludes this paper with final remarks.

\section{Original Datasets and Neural Networks} 
\label{Original Data}
This study relies on (i) mean-field simulations and initial neural network training done by Friemann et al. \cite{Friemann2023}, (ii) following transfer learning to full-field simulations by Cheung and Mirkhalaf \cite{Cheung2024_transfer}, and (iii) data augmentation method developed by Cheung et al. \cite{Cheung2024}. In these studies micro-mechanical simulations of SFRCs with specific properties for the matrix and fiber materials were conducted. Micro-mechanical properties included a variety of fiber orientations and fiber volume fractions. SFRCs were computationally modeled with 6-dimensional-strain paths randomly generated to capture their path-dependent non-linear elasto-plastic behavior.

\begin{figure}[ht!]
    \centering
    \includegraphics[scale=0.7, trim={0 0 0 0},clip]{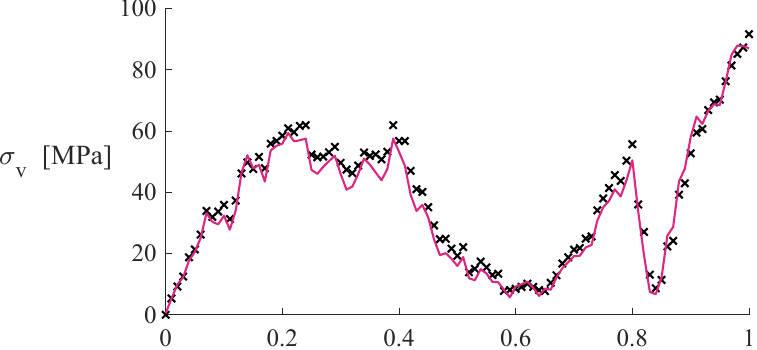}
    \includegraphics[scale=0.7, trim={0 0 0 0},clip]{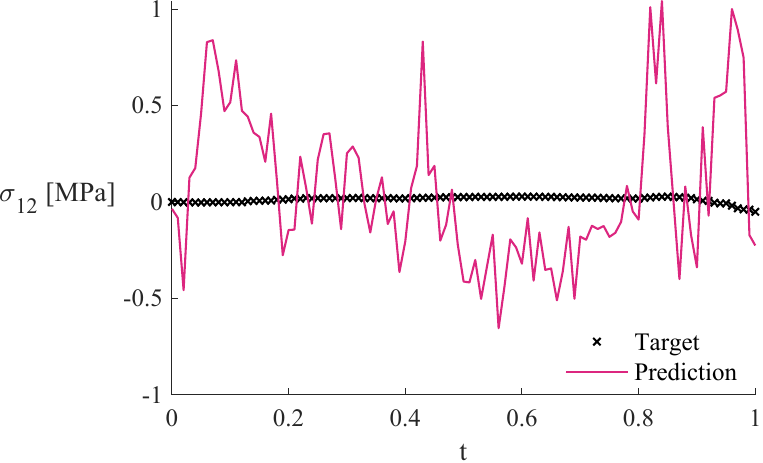}
    \caption{Micro-mechanical simulation (target) and RNN results (prediction) of von Mises stress ($\sigma_{\text{v}}$) and a near-zero shear stress component ($\sigma_{12}$).}
    \label{Uniaxial}
\end{figure}

Although accurate predictions of stress evolutions were obtained using the developed networks, in some particular loading cases, when some stress components are close to zero, a noisy prediction can be observed. Figure \ref{Uniaxial} illustrates an examples of the resulting RNN prediction, in which a shear stress component ($\sigma_{12}$) is near zero. 
The trend of a noisy signal near zero was observed in multiple occasions. The aim of this study is to quantify the variability of the error using TTA and reduce it to improve the prediction and shape accuracy of the RNN results, without a need for additional simulations or training. The following three subsections provide detailed explanations of the mean-field dataset generation and the initial neural network training (Section \hyperlink{subsection.2.1}{2.1}), the full-field data generation and the transfer learning of the RNN (Section \hyperlink{subsection.2.2}{2.2}), and the previously developed data augmentation method (Section \hyperlink{subsection.2.3}{2.3}).

\subsection{Mean-field data and network}
In a previous study by Friemann et al. \cite{Friemann2023}, mean-field simulations were conducted, and an RNN architecture was developed and implemented to model the generated data. The following steps were used for the micro-mechanical simulations and implementation of the neural network.

\subsubsection{Material model parameters}
The material constituents were chosen with a polymer matrix similar to Polyamide 6.6, with a reinforcement of short glass fibers. A fiber is represented by a unit vector \(\boldsymbol{p}\) in a Cartesian coordinate system with the axes \(x_1, x_2, x_3\). It is described by two Euler angles: $\theta$ (between \(\boldsymbol{p}\) and the axis \(x_3\)), and $\phi$ (between \(\boldsymbol{p}\) and the plane of \(x_1x_2\)) \cite{Advani1987}. The components of \(\boldsymbol{p}\) are given by
\begin{equation} \label{prob_eqn}
\boldsymbol{p}=\begin{bmatrix}
    \sin{\theta} \cos{\phi} \\
    \sin{\theta} \sin{\phi} \\
    \cos{\theta}
\end{bmatrix}.
\end{equation}
The orientation distribution function $\psi(\boldsymbol{p})$ has the following properties:
\begin{equation} \label{psi_eqn_prob}
    \psi(\boldsymbol{p}) = \psi(-\boldsymbol{p}),
\end{equation}
\begin{equation} \label{psi_eqn_integral}
    \oint \psi(\boldsymbol{p}) \,d\boldsymbol{p} = 1.
\end{equation}
The components of the second order orientation tensor (\(\boldsymbol{a}\)) is given by:
\begin{equation}
    a_{ij} = \int_{\Omega} p_i p_j \psi(\boldsymbol{p})d\boldsymbol{p},
\end{equation}
where \(\Omega\) represents the whole domain of the RVE. For the random samples, a three-dimensional orientation tensor was randomly generated. Initially diagonal orientation tensors were randomly sampled. Subsequently, the orientation tensor was subjected to a random rotation to achieve the final orientation tensor, by applying a randomly generated rotation tensor using Arvo's algorithm \cite{Arvo1992}. Following this, a corresponding volume fraction was randomly chosen between 10\% to 15\%. Following this, a corresponding volume fraction was randomly chosen between 10\% to 15\%. 

\subsubsection{Strain path generation}
Stochastic 6-dimensional strain trajectories were generated by randomly sampling and combining 6-dimensional drift and noise vectors. Then, the resulting strain path was scaled so that the maximum strain equaled the specified maximum strain. 
More detailed information about the strain path generation and material model parameters can be found in \cite{Friemann2023}.

\subsubsection{Mean-field analyses}
Using the material constituents, and randomly generating 40,000 orientation tensors and corresponding strain paths, mean field-analyses were preformed. DIGIMAT-MF was used to perform the simulations. The DIGIMAT mean-field module uses a two step homogenization method by dividing the samples into pseudo-grains (PGs) \cite{DOGHRI2011352}, and using a Hill-type incremental formulation followed by the Mori-Tanaka method \cite{Hill1965, Mori1973}.

\subsubsection{Neural Network Model Architecture}
\begin{figure}[ht!]
    \centering
    \includegraphics[scale=0.7, trim={0cm 0 0cm 0},clip]{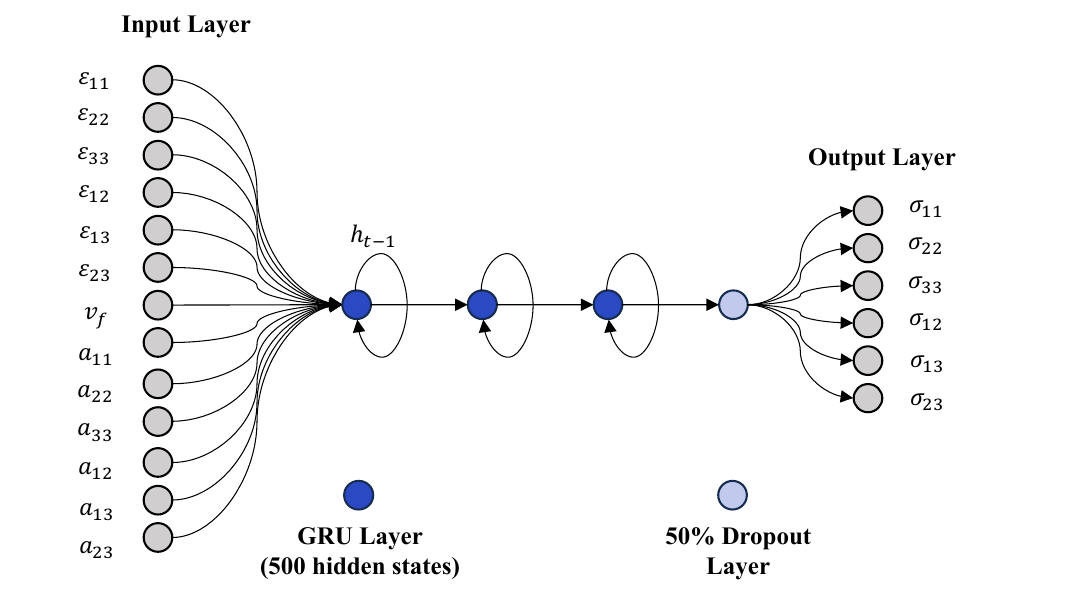}
    \caption{The RNN architecture, consisting of 3 hidden GRU layers each containing 500 hidden states, where ${h_{t}}$ is the hidden layer vector.}
    \label{RNN}
\end{figure}

The RNN developed by Friemann et al. \cite{Friemann2023} comprised 13 inputs, consisting of 6 orientation tensor components, a sequence of 6 strain tensor components, and a fiber volume fraction. The RNN architecture was constructed with three Gated Recurrent Unit (GRU) layers \cite{Cho2014}, each comprising 500 hidden states. The GRU mechanism updated the model's state for the subsequent time input. After the GRU layers, a dropout layer \cite{Srivastava2014} with a 50\% dropout rate was incorporated. The final layer contains 6 neurons, representing the 6 output stress components, thereby providing a structure capable of modeling of intricate 6-dimensional stress-strain evolutions. An illustration of the RNN architecture is presented in Figure \ref{RNN}.

It is worth noting that in this study, time step in fact refers to pseudo time step since there is no actual time dependency in the path-dependent elasto-plastic simulations.

\subsubsection{Training of the neural network}
The dataset was partitioned into training, validation, and testing data-subsets, consisting of 80\%, 19.75\%, and 0.25\% of the data, respectively. Subsequently, the RNN model was trained and validated using the corresponding datasets. 
To optimize the neural network training process, the default Matlab loss function for time-series regression was employed during the training process. This function incorporates parameters such as sequence length ($S$), number of outputs ($R$), target ($T$), and network prediction ($O$):
\begin{equation}
    \text{{loss}} = \frac{1}{2S} \sum_{i=1}^{S} \sum_{j=1}^{R} (T_{ij} - O_{ij})^2.
\end{equation}
The ADAM optimizer was used to minimize the loss function, with default parameter values selected within an appropriate range for training of the RNN \cite{Kingma2017}. The hyperparameters, \ie maximum epochs, minimum batch size, initial learning rate, learning rate drop period and factor, and gradient threshold, were adjusted based on learning rate decay relative to the number of iterations. Bayesian optimization function \cite{Snoek2012} within Matlab was utilized for hyperparameter optimization, allowing for up to 32 trials with the goal of minimizing validation loss. The final optimized parameters were determined by selecting the iteration with the lowest validation loss from the most successful trial. Moreover, to mitigate overfitting and exploding gradients, L2-regularization and gradient clipping techniques were integrated \cite{Ying2019, Pascanu2013}.

\subsection{Full-field data and transfer learning}
Cheung and Mirkhalaf \cite{Cheung2024} utilized transfer learning to fine tune the previous network developed by Friemann et al. \cite{Friemann2023}. Initially, 547 full-field data samples were generated, each comprising 100 time steps. To generate FE/FFT simulations using DIGMAT-FE, orientation tensor, volume fraction and random strain paths were generated using the methods developed in \cite{Friemann2023}. The maximum strain of the paths was randomly chosen between 0.1 and 0.5. The following subsections outline the procedure to determine the RVE size, implementing transfer learning technique, and specific loading cases tested.

\subsubsection{RVE size determination and FE/FFT-simulations}
The size of an RVE, as determined by Cheung and Mirkhalaf \cite{Cheung2024}, was chosen to have a sufficient microstructural information necessary for capturing the non-linear elasto-plastic response effectively. This was determined following the criteria outlined by Mirkhalaf et al. \cite{Mirkhalaf2016}, which stipulates that the coefficient of variation in deformation behavior should be below a predefined threshold, and the average responses should fall within an acceptable margin of error. Following the determination of RVE size, FE and FFT analyses were conducted using DIGIMAT-FE software.

\subsubsection{Transfer-learning neural network training}
A total number of 547 simulations were conducted, and the dataset was then divided into 80\% training data, 15\% validation data, and 5\% testing data. The transfer learning approach uses the same neural network architecture and training approach as in Friemann et al. \cite{Friemann2023}, however, the training process does not start from scratch and continues on the previously trained network on mean-field data. More information about the transfer learning approach can be found in \cite{Cheung2024}.

\subsubsection{Specific loading cases}
In addition to the random 6-dimensional loading data, specific loading tests were simulated to evaluate the effectiveness of the trained RNN on standard loading conditions. These tests involved cyclic loading with strain components ranging from 0 to 0.035, then to -0.035, and returning to 0. The loading cases included uniaxial normal stress ($\sigma_{11}$), uniaxial shear stress ($\sigma_{12}$), biaxial stress in two normal directions ($\sigma_{11} + \sigma_{22}$), biaxial stress in normal and shear ($\sigma_{11} + \sigma_{23}$), and 3D normal stress ($\sigma_{11} + \sigma_{22} + \sigma_{33}$). Each loading test was applied to 11 different RVEs with random orientation tensors.

\subsection{Full-field data and data augmentation}
In a previous study by Cheung et al. \cite{Cheung2024}, a method was developed to expand a limited dataset of full-field simulations of SFRCs. In comparison to transfer-learning, a data augmentation approach facilitates developing deep learning models in cases where only \textit{one dataset} exists, in this case a full-field dataset. The method involves rotating the input and output second order tensors into multiple configurations using randomly generated rotation tensors. The full-field training dataset, including the orientation tensor, strain path and stress evolution, was augmented by using fast random rotations based on the Arvo’s \cite{Arvo1992} algorithm. The random rotation tensor ($\textbf{\textit{R}}$) was applied to the strain, orientation, and subsequently to the predicted stress tensors:
\begin{equation}
    \begin{Bmatrix} \boldsymbol{a}_{r}\\ \boldsymbol{\varepsilon}_{r}\\ \boldsymbol{\sigma}_{r}\\ \end{Bmatrix} = \boldsymbol{R} \cdot \begin{Bmatrix} \boldsymbol{a}\\  \boldsymbol{\varepsilon}\\ \boldsymbol{\sigma}\\ \end{Bmatrix} \cdot \boldsymbol{R}^{T},
\end{equation}
where, \(\boldsymbol{a}\) represents the orientation tensor, \(\boldsymbol{\varepsilon}\) represents the strain tensor, and \(\boldsymbol{\sigma}\) represents the stress tensor. The subscript \(r\) denotes the rotated version of each respective tensor after the transformation by the rotation tensor \(\boldsymbol{R}\). This was used to expand the original dataset to facilitate training of an accurate RNN model with the small full-field dataset without any extra simulation or another dataset. 

\section{Test Time Data Augmentation}
\label{Test Time Data Augmentation}
In this study, we propose a TTA method following the data augmentation method we previously proposed to augment a training dataset \cite{Cheung2024}. The previous method is adapted to rotate an input (from a test dataset) to generate a prediction during inference, after which we revert the RNN outputs to the configuration of the original input, by applying the inverse rotation tensor. Effectively, the coordinate frame of the test data is randomly rotated in a 3-dimensional space, and the prediction is rotated back to the original coordinate system.


\subsection{TTA method}
\begin{figure*}[ht!]
    \centering
    \includegraphics[scale=0.7, trim={0 0 0cm 0},clip]{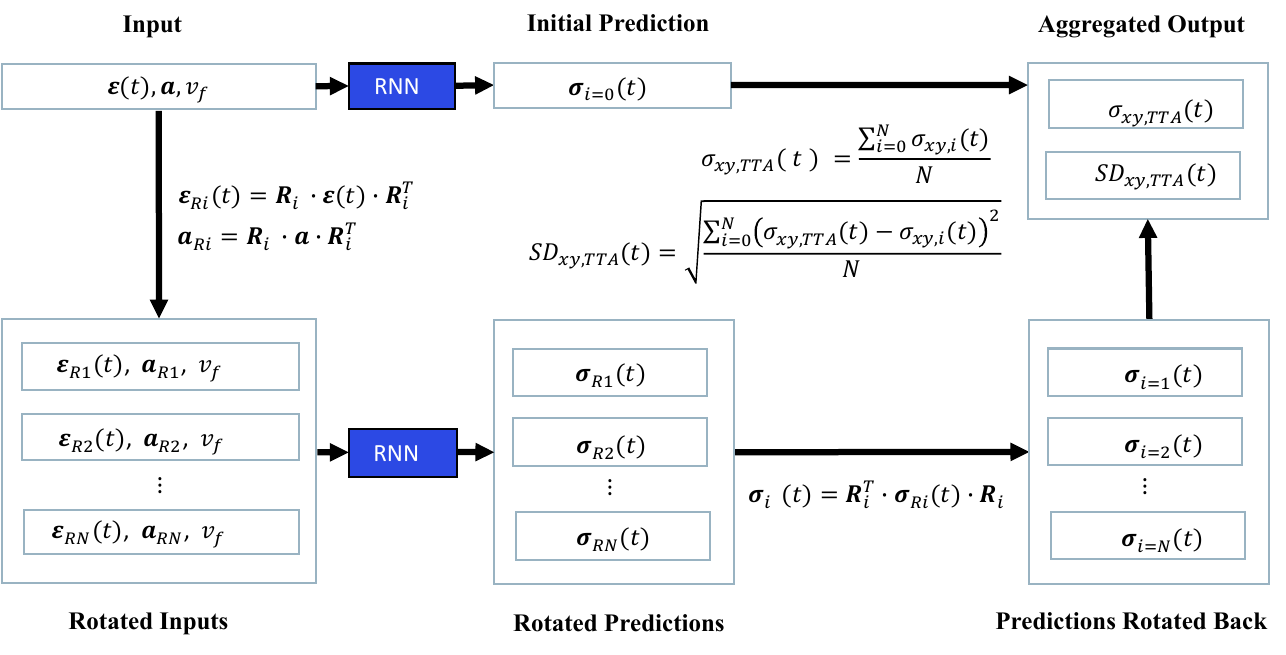} 
    \caption{A representation of the proposed TTA method, in which an input is rotated a certain number of times, and the corresponding output is rotated back to the original configuration.}
    \label{Rerotations_method}
\end{figure*}

An illustration of the proposed TTA method is presented in Figure \ref{Rerotations_method}.
In the TTA method we use Arvo's algorithm to generate a list of random rotation tensors ($\textbf{R}_i$), for $i$ from $1$ to $N$, where N is the total number of rotations. The list is extended to $i = 0$, in which the second order identity tensor is used instead of a rotation tensor, representing the initial prediction without any rotation. The list of random rotations is then used to rotate the input and rotate back the predictions made by the RNN. 

First, the orientation tensor and strain tensor are rotated, while leaving the fiber volume fraction ($\text{v}_f$) unchanged:
\begin{equation}
    \boldsymbol{a}_{Ri} = \boldsymbol{R}_{i} \cdot \boldsymbol{a} \cdot \boldsymbol{R}_{i}^{T},
\end{equation}
\begin{equation}
    \boldsymbol{\varepsilon}_{Ri}(t) = \boldsymbol{R}_{i} \cdot \boldsymbol{\varepsilon}(t) \cdot \boldsymbol{R}_{i}^{T},
\end{equation}
where the rotation is preformed for each discrete pseudo time step of \(t\), \(i\) is the rotation index changing from 0 to \(N\) (total number of rotations), and \(\boldsymbol{a}_{Ri}\) and \(\boldsymbol{\varepsilon}_{Ri}(t)\) represents the rotated orientation and strain tensors for a rotation tensor ($\textbf{R}_i$), respectively. Subsequently, the rotated inputs are introduced to the RNN model and the stress output is obtained, where $\textit{\textbf{f}(\textbf{x})}$ represents the function performed by the neural network for an input vector:
\begin{equation}
    \textit{\textbf{f}}(\boldsymbol{a}_{Ri}, \text{v}_f, \boldsymbol{\varepsilon}_{Ri}(t)) = \boldsymbol{\sigma}_{Ri}(t).
\end{equation}
Then, the calculated stress output is rotated back to the initial configuration by applying the inverse of the initial applied rotation tensor:
\begin{equation}
    \boldsymbol{\sigma}_{i}(t) = \boldsymbol{R}_{i}^T \cdot \boldsymbol{\sigma}_{Ri}(t) \cdot \boldsymbol{R}_{i}.
\end{equation}
Using the explained method, a number of predicted stress tensors ($\boldsymbol{\sigma}_{i}(t)$) are obtained for a given input of the test dataset. Each stress tensor contains 6 stress components (\(\sigma_{11,i}(t)\), \(\sigma_{22,i}(t)\), \(\sigma_{33,i}(t)\), \(\sigma_{12,i}(t)\), \(\sigma_{13,i}(t)\), \(\sigma_{23,i}(t)\)). To find the aggregated stress path, denoted \(\sigma_{xy,TTA}(t)\), where \(xy\) represents the stress component, we calculate the mean stress at each time point of the back-rotated predictions, using the following equation:
\begin{equation}
\label{aggregated_path_eq}
\sigma_{xy,TTA}(t) = \frac{\sum_{i=0}^{N} \sigma_{xy,i}(t)}{N}.
\end{equation}
The standard deviation at each time step \(SD_{xy,TTA}(t)\), is computed using all the predictions rotated back:
\begin{equation}
     SD_{xy,TTA}(t)= \sqrt{\frac{\sum_{i=1}^{N} (\sigma_{xy,i}(t) - \sigma_{xy,TTA}(t))^2}{N}}.
\end{equation}
The von Mises stress of the target (micro-mechanical simulations) is defined by:
\begin{equation}
\label{von_mises}
    \resizebox{0.9\hsize}{!}{$
    \sigma_{\text{v}}(t) = \sqrt{\frac{1}{2}\left[(\sigma_{11}(t) - \sigma_{22}(t))^2 + (\sigma_{22}(t) - \sigma_{33}(t))^2 + (\sigma_{33}(t) - \sigma_{11}(t))^2\right] + 3(\sigma_{12}^2(t) + \sigma_{13}^2(t) + \sigma_{23}^2(t))}
    $},
\end{equation}
where all six components of stress are taking part in the definition, and thus, it can be considered as a good choice for measuring a network accuracy for stress predictions. We also calculate the von Mises stres of the aggregated stress path (from the TTA method) which is denoted \(\sigma_{\text{v}, TTA}(t)\), and von Mises stress of individual predictions, denoted by \(\sigma_{\text{v},i}(t)\).
%
%
%
The standard deviation of the aggregated von Mises stress is then calculated by: 
\begin{equation}
\label{SD_v_TTA}
     SD_{\text{v}, TTA}(t)= \sqrt{\frac{1}{N} \sum_{i=1}^{N} (\sigma_{\text{v},i}(t) - \sigma_{\text{v},TTA}(t))^2}.
\end{equation}
Using the TTA method, it is possible to investigate prediction variation from the model, with the aim to enhance the robustness of RNN's predictions capabilities and provide an uncertainty estimate through averaging the predictions.

\subsection{Evaluation metrics}
The performance of the trained RNN is assessed based on its ability to predict the von Mises stress. From this, two key evaluation metrics, namely the Mean Relative Error (MeRE) and the Maximum Relative Error (MaRE), are calculated as follows:
\begin{equation}
\label{MeRE_i}
\text{MeRE}_i = \frac{1}{M} \sum_{m=1}^M \frac{{\sqrt{\sum_{t=1}^{T} ( \sigma_{\text{v}}(t) - \sigma_{\text{v},i}(t))^2}}}{{\max(\sigma_{\text{v}}(t))T}},
\end{equation}
\begin{equation}
\label{MaRE}
\text{MaRE}_i = \frac{1}{M} \sum_{m=1}^M \frac{{\max(\sigma_{\text{v}}(t) - \sigma_{\text{v},i}(t))}}{{\max(\sigma_{\text{v}}(t))}}.
\end{equation}
In which, the predicted (\(\sigma_{\text{v},i}(t)\)) and target (\(\sigma_{\text{v}}(t)\)) von Mises stress are values over the pseudo time increments (\(t=1,...,T\)). The evaluation metrics are taken as mean over the whole dataset: \(m\) represents each data sample in the test dataset (\(m=1,...,M\)) where \(M\) is the total number of data samples in the test dataset. For a specific rotation tensor (\(\textbf{R}_i\)), the corresponding MeRE of the output stress is denoted by \(\text{MeRE}_i\), and the prediction of the original configuration is denoted by \(\text{MeRE}_{i=0}\). To evaluate the variation in MeRE across all the rotations, the average MeRE (\(\text{MeRE}_{av}\)) of back-rotated predictions is calculated as follows:
\begin{equation}
\label{MeRE_AVG}
\text{MeRE}_{av} = \frac{\sum_{i=0}^{N}\text{MeRE}_i}{N+1},
\end{equation}
where \(N\) represents the total number of rotations tensors.
The MeRE and MaRE of the aggregated von Mises path (\(\text{MeRE}_{TTA}\), \(\text{MaRE}_{TTA}\)) are calculated as follows:
\begin{equation}
\label{MeRE_TTA}
\text{MeRE}_{TTA} = \frac{1}{M} \sum_{m=1}^M \frac{{\sqrt{\sum_{t=1}^{T} ( \sigma_{\text{v}}(t) - \sigma_{\text{v},TTA}(t))^2}}}{{\max(\sigma_{\text{v}}(t))T}},
\end{equation}
\begin{equation}
\label{MaRE_TTA}
\text{MaRE}_{TTA} = \frac{1}{M} \sum_{m=1}^M \frac{{\max(\sigma_{\text{v}}(t) - \sigma_{\text{v},TTA}(t))}}{{\max(\sigma_{\text{v}}(t))}}.
\end{equation}
\section{Results and Discussion} 
\label{Results}
In this section, we present the results obtained from the proposed TTA method: first, we analyze the variation in RNN predictions by mapping the \(\text{MeRE}_i\) for 100,000 rotated predictions. Next, we compare the aggregated prediction and the original prediction with the target of the simulations. Finally, we analyze standard deviation of  the aggregated von Mises path with respect to prediction error.

\subsection{Distribution of prediction variation}
\begin{figure}[ht]
    \centering
    \includegraphics[scale=0.7, trim={0 0 0 0.5cm},clip]{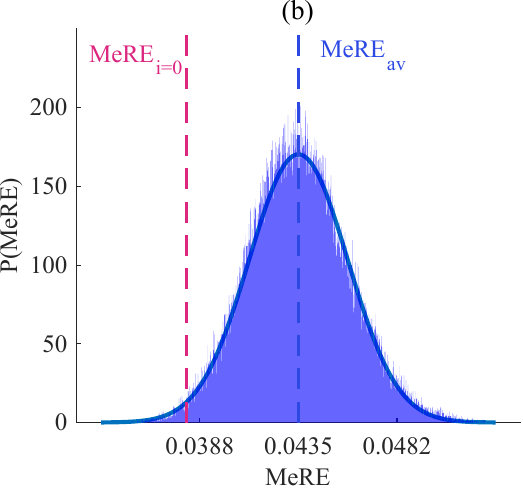}
 \caption{The corresponding probability distribution of \(\text{MeRE}_i\) values for 100,000 back-rotated predictions.}
    \label{normal_dist_fig}
\end{figure}

Provided with 100,000 back-rotated predictions, we calculate the corresponding \(\text{MeRE}_i\) for the von Mises stress (Equation (\ref{MeRE_i})) for each prediction. Subsequently, the prediction variability of \(\text{MeRE}_i\) is analyzed. The \(\text{MeRE}_{av}\) is calculated as given in Equation (\ref{MeRE_AVG}), and the standard deviation is calculated by
\begin{equation}
     \text{SD}_{MeRE}= \sqrt{\frac{1}{N} \sum_{i=1}^{N} (\text{MeRE}_{i} - \text{MeRE}_{av})^2}.
\end{equation}
We fit the probability distribution of \(\text{MeRE}_i\) to a normal distribution function. Figure \ref{normal_dist_fig} illustrates the \(\text{MeRE}_i\) distribution for each random rotation tensor ($\textbf{R}_{i}$), from a total of 100,000 individual back-rotated predictions. The histogram data have a bin width of 0.00001, where the area represents the probability density of the corresponding \(\text{MeRE}_i\) value being contained within the bounds of the bin.
For the test dataset, $\text{MeRE}_{av} = 0.0435$ with a standard deviation of $\text{SD}_{MeRE}$ = 0.0023. The probability density function (PDF), depicting the data distribution of $\text{MeRE}_{i}$, is presented, with the x-axis indicating two standard deviations per step on either side of the $\text{MeRE}_{av}$. Notably, the MeRE of the initial coordinate system, $\text{MeRE}_{i=0}$, has a value of 0.0382, placing it in the 90\textsuperscript{th} percentile and being in the upper margin compared to the back-rotated predictions. However, the accurate initial prediction is a particular case for the provided test dataset and the specific initial configuration, while alternative configurations, and their corresponding coordinate systems (or other datasets) could result in less accurate initial predictions. Hence, since each rotation of the coordinate system remains physically meaningful, each rotated simulation is equally valid. Consequently, the $\text{MeRE}_{av} = 0.0435$ offers a more accurate assessment of the network's performance compared to the initial prediction. 

\begin{figure*}[ht]
\centering
\includegraphics[scale=0.7, trim={0 1 0 1},clip]{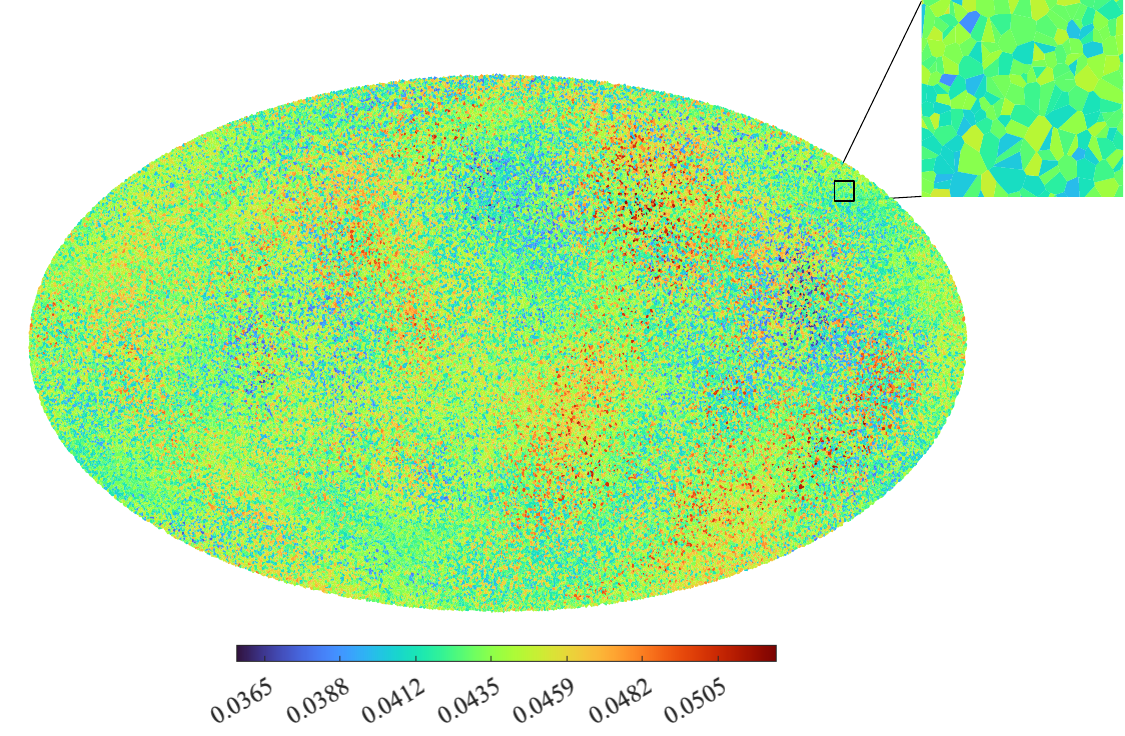}
\caption{\(\text{MeRE}_i\) of each corresponding back-rotated prediction on a unit sphere, projected on a 2D plane using Mollweide projection \cite{Feeman2000}.}
\label{voronoi_mollweide}
\end{figure*}

To further analyze how MeRE varies in relation to the rotation tensors, we map the $\text{MeRE}_{i}$ value of each rotation onto a spherical coordinate system. Each back-rotated prediction, \(\boldsymbol{\sigma}_i\), has a corresponding rotation tensor ($\textbf{R}_{i}$). Therefore, by initiating from the point [0 0 1] and applying the rotation tensor and assigning the $\text{MeRE}_{i}$ value to the corresponding point, we can visualize the variation of MeRE in relation to the coordinate system. The resulting sphere is then projected onto a 2D surface using a Mollweide projection \cite{Feeman2000}. The Mollweide projection is a pseudo-cylindrical map projection used for representing the entire surface of a sphere. To perform the Mollweide projection, the following calculations are needed.
First, we define the latitude \(\varphi\) and longitude \(\lambda\) from the Cartesian coordinates \((x, y, z)\):
\begin{equation}
\varphi = \sin^{-1}\left(\frac{z}{\sqrt{x^2 + y^2 + z^2}}\right),
\end{equation}
\begin{equation}
\lambda = \tan^{-1}\left(\frac{y}{x}\right).
\end{equation}
The relationship between the auxiliary angle \(\theta\) and the latitude \(\varphi\) is given by
\begin{equation}
2\theta + \sin(2\theta) = \pi \sin(\varphi).
\end{equation}
Finally, the x and y coordinates in the Mollweide projection are calculated using the following equations:
\begin{equation}
x = R \frac{2\sqrt{2}}{\pi} \lambda \cos(\theta),
\end{equation}
\begin{equation}
y = R \sqrt{2} \sin(\theta),
\end{equation}
where the radius $R$ is equal to 2. These equations transform the spherical coordinates into a 2D plane using the Mollweide projection, allowing for a representation of the entire surface of the sphere. The points of MeRE are then interpolated using the Voronoi method, to improve visualization by avoiding empty space and overlapping points \cite{Aurenhammer1991}. The Voronoi method partitions the plane into regions, or Voronoi cells, based on the distance to a given set of points. Each Voronoi cell contains all the points that are closer to its corresponding seed point than to any other seed point. Through this technique, the MeRE for each prediction is visualized, as illustrated in Figure \ref{voronoi_mollweide}.
The plot reveals clusters of both high and low MeRE values, with adjacent points on the sphere exhibiting similar trends. Certain areas display more homogeneous clusters, while other areas are more noisy.

\subsection{Numerical error}
To ensure that the variation we observe in rotated predictions is not a result of rounding errors introduced by the rotation algorithm, a numerical analysis of the algorithm is performed. Specifically, we subject the input, the target output, and the network stress prediction, to rotation tensors and subsequent inverse rotations. The maximum absolute value of the difference, on average across simulations, is calculated by:
\begin{equation}
    \text{\textit{numerical error}} = \frac{1}{M} \sum_{m=1}^{M} \max_{t} \left( \left| \mathbf{x}_m(t) - \mathbf{R}^T \cdot \left( \mathbf{R} \cdot \mathbf{x}_m(t) \cdot \mathbf{R}^T \right) \cdot \mathbf{R} \right| \right),
\end{equation}
where \textbf{x}, represents the augmented tensors, \ie the neural network input tensors (containing strain and orientation tensor), the target stress tensor, and the output stress tensor. The resulting numerical computational errors incurred during this process are quantified and presented in Table \ref{tab:numerical_error}. 
\begin{table}[ht]
    \caption{Numerical computing error of rotating and rotating back the input, output, and the target prediction, averaged over the entire test dataset.}
    
    \label{tab:numerical_error}
    \centering
    \scalebox{0.8}{
    \begin{tabular}{c c c}
        \toprule
        \textbf{Input} & \textbf{Target} & \textbf{Output} \\
        \midrule
        $1.9487 \times 10^{-15}$ & $6.7719 \times 10^{-14}$ & $7.2557 \times 10^{-14}$ \\
        \bottomrule
    \end{tabular}
    }
\end{table}
Numerical errors are on the order of $10^{-14}$, confirming that rounding and computational approximations have almost no impact on the results. This high level of precision suggests that the rotation algorithm is robust, ensuring that any variations observed in predictions are not artifacts of numerical noise but rather reflect actual differences in the model predictions.

\subsection{TTA aggregation for time-signal noise-reduction}
\begin{figure}[ht]
    \centering
    \includegraphics[scale=0.7, trim={0 0 0 0}, clip]{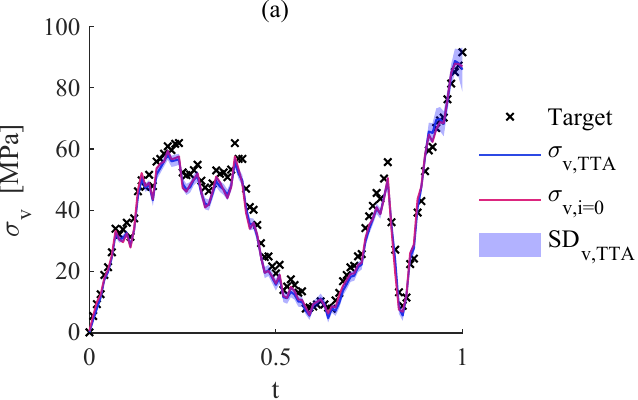}
    \includegraphics[scale=0.7, trim={0 0 0 0}, clip]{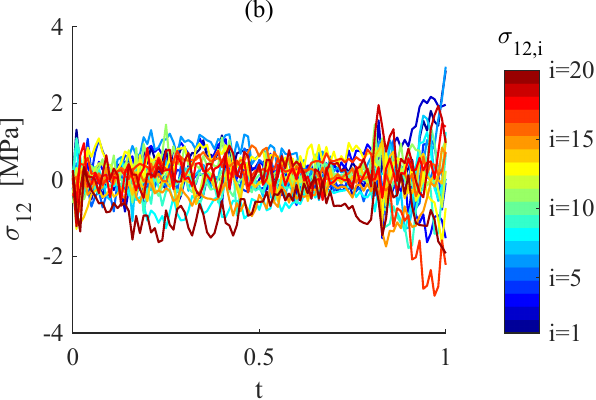}
    \includegraphics[scale=0.7, trim={0 0 0 0}, clip]{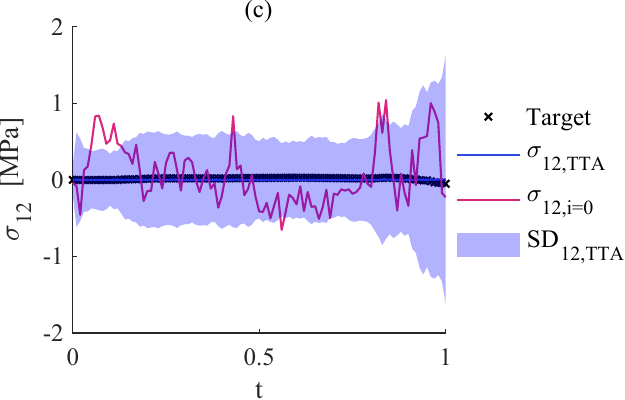}
    \includegraphics[scale=0.7, trim={0 0 0 0}, clip]{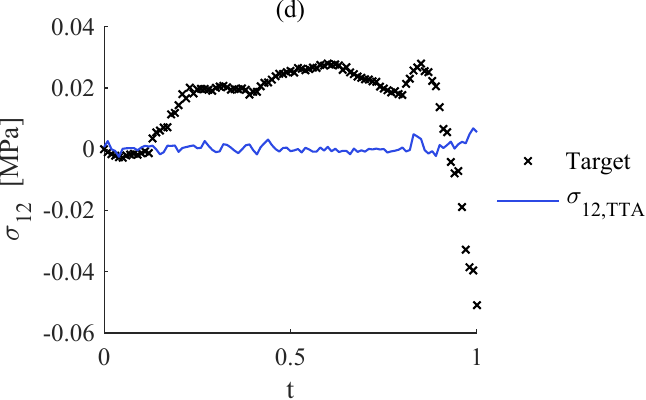}
    \caption{An example plot of (a) von Mises stress, showing the aggregated path of ($\sigma_{\text{v},TTA}(t)$) compared with the initial prediction and simulation target, (b) The first 20 back-rotated predictions of shear stress components of $\sigma_{12, i}(t)$, (c) The standard deviation of the TTA method for shear stress component ($\sigma_{12}(t)$), (d) Target shear stress component $\sigma_{12}(t)$ compared with $\sigma_{12, TTA}(t)$. }
    \label{cell6}
\end{figure}

The aggregation of back-rotated predictions can compensate inaccurate individual predictions, by combining all the predicted paths. Figure \ref{cell6} shows plots comparing the von Mises and $\sigma_{12}$ stress of a simulation with the target, initial predicted path, aggregated path obtained using the TTA method, and the corresponding standard deviation of back-rotated predictions. Additionally, it shows predictions rotated back to the original coordinate system for stress component \(\sigma_{12,i}(t)\), where i ranges from 1 to 20. It also includes the standard deviation of the corresponding prediction shown by the shaded area. The bounds show a variation in predictions from the RNN. The aggregated path derived from multiple predictions rotated back shows a considerable improvement compared with a single prediction generated by the RNN.

\begin{figure}[ht]
    \centering
    \includegraphics[scale=0.7, trim={0 0 0 0},clip]{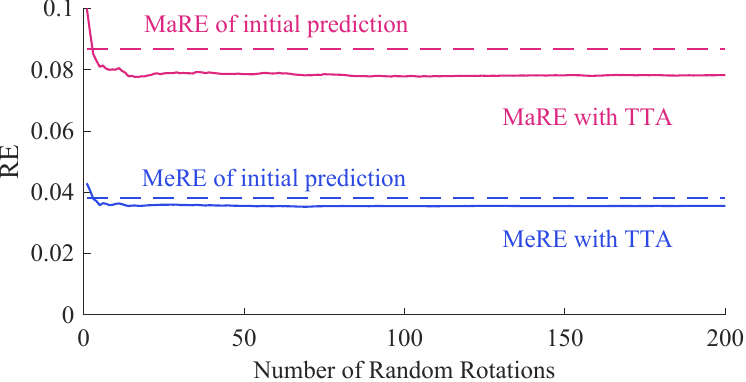}
 \caption{\(\text{MeRE}_{i=0}\) and \(\text{MaRE}_{i=0}\) compared with \(\text{MeRE}_{TTA}\) and \(\text{MaRE}_{TTA}\)  for different number of random rotations.}
    \label{MeRE_SRi}
\end{figure}

To evaluate the impact of the number of rotations used in the TTA approach, the \(\text{MeRE}_{TTA}\) and \(\text{MaRE}_{TTA}\) are calculated for different number of random rotations. The number of rotations used in the TTA method ranges from 1 to 200, and the results are presented in Figure \ref{MeRE_SRi}. 
Specifically, the average path resulting from 200 random rotations exhibits an \(\text{MeRE}_{TTA}\) of 0.0356, whereas the \(\text{MeRE}_{av}\), the average MeRE of single predictions is 0.0435. Thus, aggregating the paths results in a reduction in MeRE of approximately 19\%. The reduction in MeRE achieved with TTA demonstrates a significant improvement in prediction accuracy. This 19\% reduction demonstrates the effectiveness of the TTA method in minimizing errors by averaging predictions across multiple rotations. In comparison, the \(\text{MeRE}_{i=0}\) for the original coordinate system has a value of 0.0382. When comparing the \(\text{MeRE}_{TTA}\) with \(\text{MeRE}_{i=0}\), the reduction in MeRE is approximately around 7\%, indicating that the prediction in original coordinate system is relatively accurate compared to predictions in alternative coordinate systems. The \(\text{MeRE}_{TTA}\) reaches a plateau around 200 predictions. Moreover, adding a total of 100,000 back-rotated predictions, yields in a minor difference of \(\text{MeRE}_{TTA}\) of 0.0358. Moreover, it should be noted that as the \(\text{MeRE}_{TTA}\) reaches a stationary phase, suggesting that \(\sigma_{\text{v},TTA}(t)\) reaches equivariance in regards to rotations. The result of the repeats compared with the  \(\text{MeRE}_{i=0}\) and \(\text{MeRE}_{TTA, N=100,000}\) are presented in Table \ref{tab:Repeats}. 
\begin{table*}[ht]
    \caption{MeRE and MaRE for different repetitions of TTA with 200 random rotations. The starting MeRE and MaRE for one prediction, and the final MeRE and MaRE are the value reached after TTA with 100,000 random rotations.}
    \centering
    \scalebox{0.8}{
    \begin{tabular}{cccccc c c}
    \hline
    \toprule
    \textbf{Repeat:} & \textbf{1} & \textbf{2} & \textbf{3} & \textbf{4} & \textbf{5} & \textbf{Initial} & \textbf{Final} \\
    \textbf{} & \textbf{N=200} & \textbf{(N=200)} & \textbf{(N=200)} & \textbf{(N=200)} & \textbf{(N=200)} & \textbf{(i=0)} & \textbf{ (N=100,000)} \\
    \midrule
    MeRE & 0.03558 & 0.03593 & 0.03590 & 0.03582 & 0.03583 & 0.03820 & 0.03579 \\
    MaRE & 0.07827 & 0.07898 & 0.07862 & 0.07918 & 0.07815 & 0.08688 & 0.07874 \\
    \bottomrule
    \hline
    \end{tabular}
    }
    \label{tab:Repeats}
\end{table*}

\begin{figure}[ht]
    \centering
    \includegraphics[scale=0.7, trim={0 0 0 0}, clip]{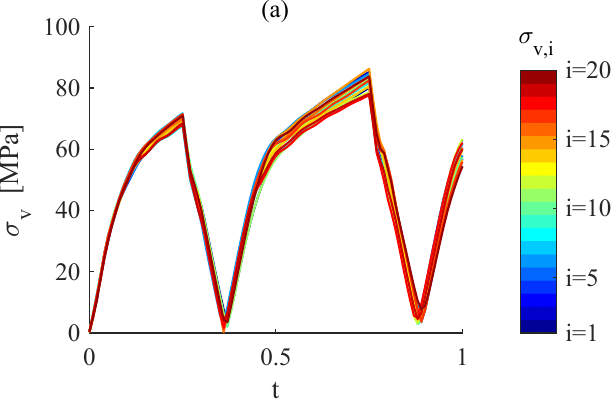} 
    \includegraphics[scale=0.7, trim={0 0 0 0}, clip]{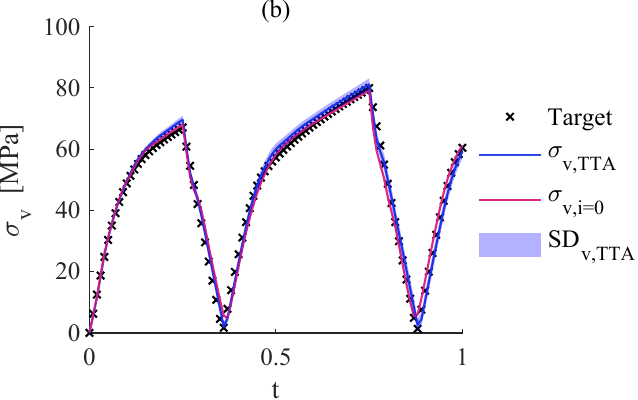}
    \includegraphics[scale=0.7, trim={0 0 0 0}, clip]{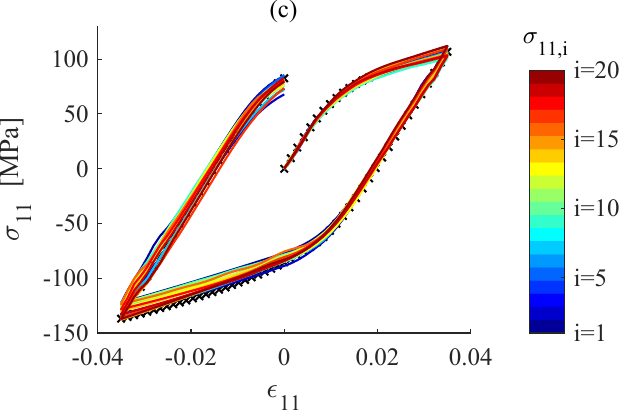}
    \includegraphics[scale=0.7, trim={0 0 0 0}, clip]{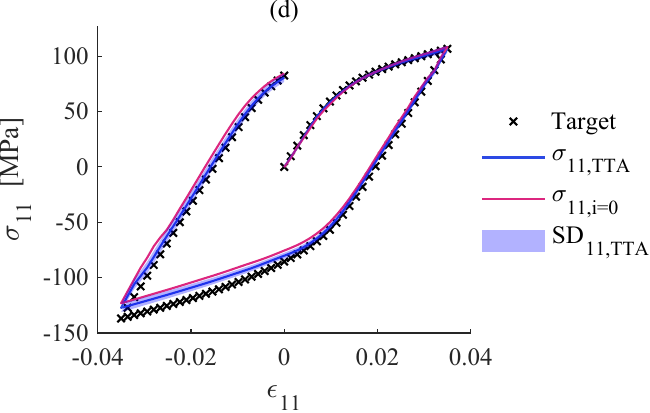}
    \caption{An example plot of uniaxial loading: (a) The first 20 back-rotated predictions of von Mises stress, (b) The aggregated path of ($\sigma_{\text{v},TTA}(t)$) compared with the initial prediction and simulation target, (c) The component ($\sigma_{\text{11},i}(t)$) versus ($\varepsilon_{\text{11},i}(t)$), (d) The TTA results of component ($\sigma_{\text{11},TTA}(t)$) in comparison to ($\varepsilon_{\text{11},i}(t)$)}
    \label{cell6_uni}
\end{figure}

\begin{figure}[ht]
    \centering
    \includegraphics[scale=0.7, trim={0 0 0 0},clip]{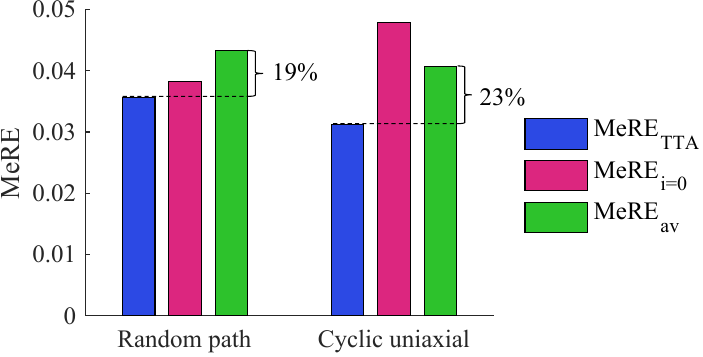}
 \caption{MeRE values of the TTA method (\(\text{MeRE}_{TTA}\)), the initial prediction (\(\text{MeRE}_{i=0}\)), and average of individually rotated predictions (\(\text{MeRE}_{av}\)) for random and uniaxial loading datasets.}
    \label{Random_vs_cyclic}
\end{figure}

To assess variations of initial predictions for other datasets, we apply the TTA method to a dataset containing specific loading conditions. In particular, a specific loading dataset containing solely uniaxial loading, developed by Cheung and Mirkhalaf \cite{Cheung2024}, where a cyclic load in \(\sigma_{11}(t)\) was applied in cycles, resulting in stress \(\sigma_{11}(t)\) while the remaining stress components are zero. An example plot of uniaxial loading case is shown in Figure \ref{cell6_uni}.
The various MeRE values are presented in Figure \ref{Random_vs_cyclic}. 
For the uniaxial test dataset, by comparing the \(\text{MaRE}_{i=0}\) with \(\text{MaRE}_{TTA}\), it decreases from 0.121 to 0.0791, representing a 45\% reduction. The MeRE, \ie \(\text{MeRE}_{i=0}\) compared with \(\text{MeRE}_{TTA}\) decreases from 0.0479 to 0.0313, or 35\%.
Out of the 11 samples of uniaxial cyclic loading, 10 samples have a lower \(\text{MeRE}_{TTA}\), compared to the \(\text{MeRE}_{i=0}\). By calculating the $\text{MeRE}_{av}$ using the approach explained in section \hyperlink{subsection.4.1}{4.1}, it has a value of 0.0407, which is lower than the \(\text{MeRE}_{i=0}\). This corresponds with an actual decrease in MeRE of 23\%, which is in good agreement with the result from the random generated loading paths of 19\%. This shows that, for the RNN used in this study, the TTA method results in an approximate improvement of accuracy of 20\%.

\subsection{Shape consistency}
\begin{figure}[ht]
    \centering
    \includegraphics[scale=0.7, trim={0cm 0cm 0cm 0cm},clip]{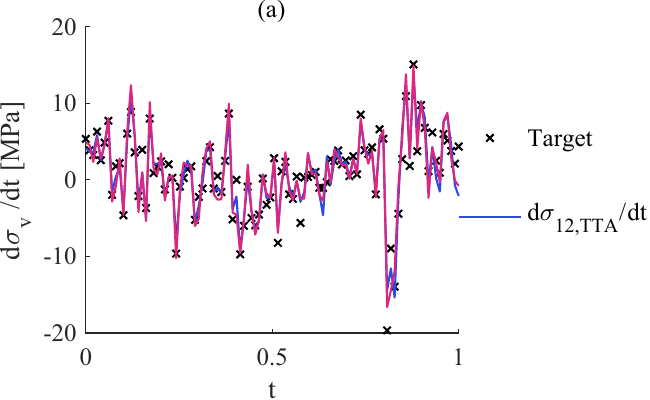}
    \includegraphics[scale=0.7, trim={0cm 0cm 0cm 0cm},clip]{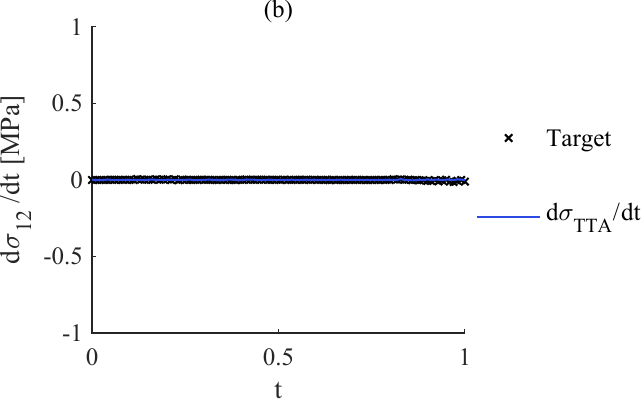}
    \includegraphics[scale=0.7, trim={0cm 0cm 0cm 0cm},clip]{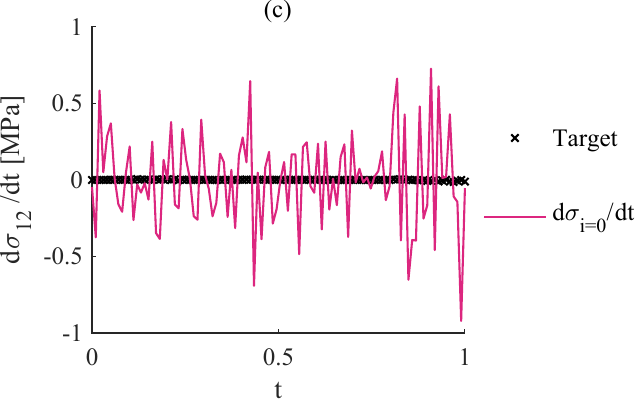}
    \caption{Example plots of the first-order derivative of stress predictions: (a) First-order derivative of the von Mises stress for the target simulation, aggregated path (\(\sigma_{\text{v},TTA}(t)\)), and initial prediction (\(\sigma_{\text{v},i=0}(t)\)), (b) First-order derivative of stress component 12, for the target simulation and aggregated path (\(\sigma_{12,TTA}(t)\)), (c) First-order derivative of stress component 12, for the target simulation and initial prediction (\(\sigma_{12,i=0}(t)\)).}
    \label{Derivative}
\end{figure}

To illustrate how the shape consistency (in stress predictions) is improved using the TTA method, an example simulation of the first-order derivative is presented in Figure \ref{Derivative}.
The shape prediction is analyzed by taking the first-order differences along the time-series. Then, by evaluating the correlation coefficient between the first-order differences of the predicted and the target path, a metric of the correlation is obtained. This approach negates the offset of the predicted path, by first taking the first-order differences followed by the correlation coefficient. In the following paragraphs we explain shape analysis and the results.

\begin{figure}[ht]
    \centering
    \includegraphics[scale=0.7, trim={0 0 0 0},clip]{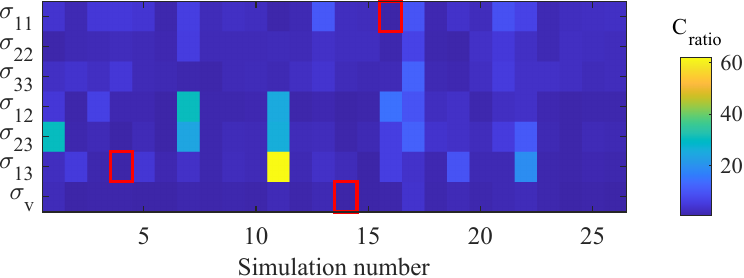}
    \caption{The ratio of correlation coefficients ($C_{ratio}$) of all stress components, plotted for each simulation in the test dataset.}
    \label{correlation_coefficients}
\end{figure}

First, we compute the Pearson correlation coefficients between each fist-order differences of the RNN predicted von Mises stress, with fist-order differences of the target simulation. The Pearson correlation coefficient ($r$) \cite{Pearson1895, Rodgers1988} is a measure of the linear correlation between a group of variables, namely $X(t)$ and $Y(t)$:
\begin{equation}
r = \frac{\sum_{t=1}^{T} (X(t) - \bar{X}) (Y(t) - \bar{Y})}{\sqrt{\sum_{t=1}^{T} (X(t) - \bar{X})^2} \sqrt{\sum_{t=1}^{T} (Y(t) - \bar{Y})^2}},
\end{equation}
where \(\bar{X}\) and \(\bar{Y}\) are the means of the sample points throughout the time series of \(X(t)\) and \(Y(t)\) respectively, and \(t\) is the index over the \(T\) time points. 
For the initial prediction for the von Mises stress (\(r_{\text{v},i=0}\)), the variables \(X(t)\) and \(Y(t)\) are defined as
\begin{equation}
X(t) = \frac{d\sigma_{\text{v},i=0}(t)}{dt}, \quad Y(t) = \frac{d\sigma_{\text{v}}(t)}{dt}.
\end{equation}
Second, we compute the correlation coefficient for the TTA aggregated path (\(r_{\text{v},TTA}\)), the variables \(X(t)\) and \(Y(t)\) are given by
\begin{equation}
X(t) = \frac{d\sigma_{\text{v},TTA}(t)}{dt}, \quad Y(t) = \frac{d\sigma_{\text{v}}(t)}{dt}.
\end{equation}
Last, the ratio of the correlation coefficients (\(C_{ratio}\)) is calculated to compare the initial prediction with the aggregated predictions using the following equation:
\begin{equation}
C_{ratio} = \frac{1 - r_{\text{v},i=0}}{1 - r_{\text{v},TTA}}.
\end{equation}
Similarly, the stress component can be changed from the von Mises stress to any component of stress tensor for the target stress, initial prediction, and aggregated stress path. 
The ratio will be greater than 1 if there is an improvement using the TTA method. Each ratio of the correlation coefficients is plotted in Figure \ref{correlation_coefficients} for all stress components and the von Mises stress, for each simulation in the test dataset.
The average correlation ratio across the dataset is 1.9737, indicating an improvement in the prediction using the TTA method. In only three cases, marked with a red rectangle in Figure \ref{correlation_coefficients}, the correlation ratio is less than 1. The analysis of first-order differences shows that the TTA method significantly improves shape consistency in predictions. This improvement in shape consistency can be interpreted as another indication of better capturing of the path-dependency in the simulations. 

\subsection{Uncertainty prediction}
\begin{figure}[ht]
    \centering
    \includegraphics[scale=0.7, trim={0cm 0cm 0cm 0cm},clip]{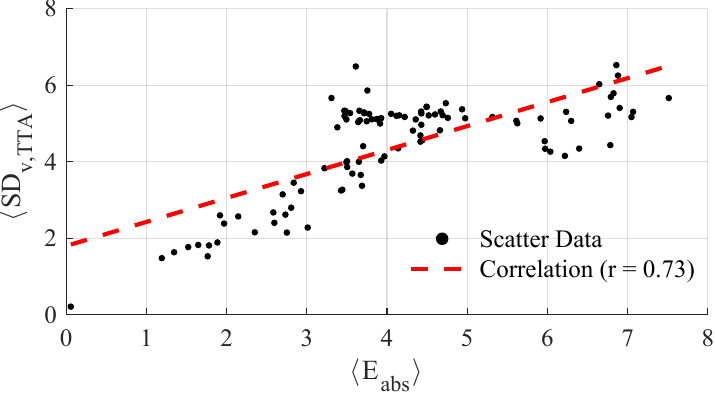}
    \caption{Scatter plot of absolute error versus average standard deviation from the TTA method.}
    \label{Scatter}
\end{figure}

To evaluate uncertainties from the TTA method, we compute the average standard deviation across the dataset for each time step, \(\langle{SD}_{\text{v},TTA}\rangle(t)\), as follows:
\begin{equation}
    \langle{SD}_{\text{v},TTA}\rangle(t)  = \frac{1}{M} \sum_{m=1}^M SD_{\text{v},TTA, m}(t),
\end{equation}
where \(SD_{\text{v},TTA,m}(t)\) is defined in Equation (\ref{SD_v_TTA}), here m is the simulation number from 1 to 26. The average prediction error is given by:
\begin{equation}
E_{abs,m}(t) = | \sigma_{\text{v},m}(t) - \sigma_{\text{v},TTA,m}(t) |,
\end{equation}
\begin{equation}
    \langle E_{abs} \rangle(t) = \frac{1}{M} \sum_{m=1}^M E_{abs, m}(t),
\end{equation}
where \(\sigma_{\text{v},m}(t)\) represents the target von Mises stress from a simulation in the dataset, and \(\sigma_{\text{v},TTA,m}\) is the predicted von Mises stress obtained from the TTA method (Equation (\ref{von_mises})). This equation quantifies the absolute deviation between the predicted and actual stress values at each pseudo time step. We evaluate at each pseudo time step, since the uncertainty of the RNN can change over time. Figure \ref{Scatter} shows a scatter plot of absolute error versus the average standard deviation.
By comparing the average SD with the average prediction error, as depicted in Figure \ref{Scatter}, we observe a linear correlation. This plot shows linear correlation with a coefficient of r = 0.73. This suggests that the TTA method provides an estimate of prediction uncertainty. The correlation coefficient of 0.73 indicates that the variance captured by TTA relates with the actual prediction errors, supporting the use of TTA as an effective tool for quantifying prediction uncertainty.

However, the absolute error versus standard deviations in individual simulations, for each stress components, has a correlation coefficient of r = 0.63, which is considerably lower than the correlation observed in Figure \ref{Scatter}. The reduced correlation indicates that the relationship between individual prediction uncertainties and actual errors is less consistent. For individual predictions, prediction errors are is inconsistent, resulting in a noisy outcomes. Therefore, while the SD correlates with individual simulations and could be used for uncertainty estimation for individual cases, the higher correlation coefficient in Figure \ref{Scatter} demonstrates that averaging provides a more reliable estimate of uncertainty for the dataset as a whole.

\begin{figure}[ht]
    \centering
    \includegraphics[scale=0.7, trim={0cm 0cm 0cm 0cm},clip]{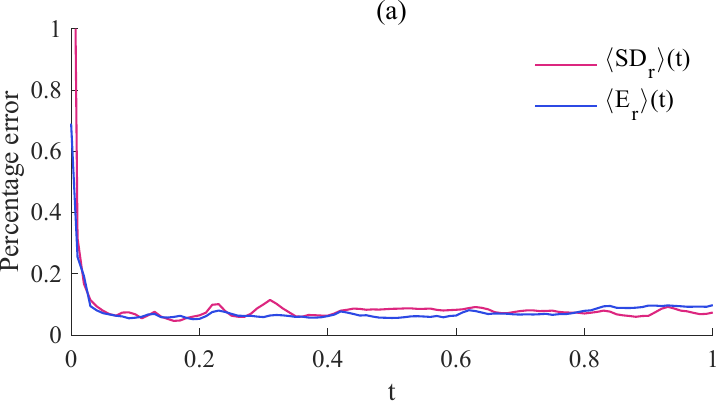}
    \includegraphics[scale=0.7, trim={0cm 0cm 0cm 0cm},clip]{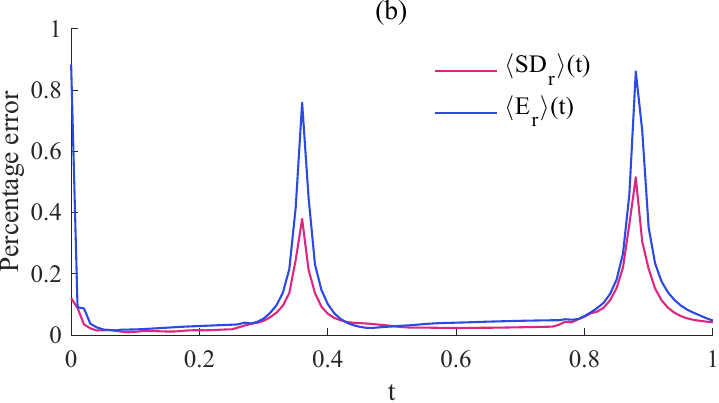}
    \caption{Comparison of normalized average standard deviation and average relative error of the predictions for (a) Random load cases, (b) Uniaxial stress tests.}
    \label{SD_vs_error_uni}
\end{figure}

To further evaluate the uncertainty estimate, we normalize the error of the prediction and the average SD by dividing them by the von Mises stress of the simulation or TTA aggregated path prediction, respectively.
The relative error is then computed by:
\begin{equation}
    \langle{{E}_{r}}\rangle(t) = \frac{1}{M} \sum_{m=1}^M \frac{E_{abs,m}(t)}{\sigma_{\text{v},TTA,m}(t)}.
\end{equation}
This equation provides a dimensionless measure of the average error relative to the magnitude of the stress.
To obtain a normalized standard deviation, \( \langle{{SD}_{r}}\rangle\), we use:
\begin{equation}
\langle{{SD}_{r}}\rangle(t) =  \frac{1}{M} \sum_{m=1}^M \frac{SD_{\text{v},TTA,m}(t)}{\sigma_{\text{v},TTA,m}(t)},
\end{equation}
where \(SD_{\text{v},TTA,m}(t)\), and \(\sigma_{\text{v},TTA,m}(t)\) are defined in Equations (\ref{SD_v_TTA}) and (\ref{von_mises}) respectively. We also calculate the average SD for the set of uniaxial loading dataset and compare this with the average relative error of the predictions. The results for the random and uniaxial dataset are presented in Figure \ref{SD_vs_error_uni}.
This normalization allows us to assess how the variability in the back-rotated predictions compares to the stress magnitude, providing a relative measure of prediction uncertainty. The normalized standard deviation correlates with the relative error with r = 0.95, thereby demonstrating that the TTA methods provides a reliable uncertainty estimate.

Overall, the TTA method proves to be a robust approach for uncertainty estimation, leveraging rotational transformations. Compared to conventional methods, such as ensemble methods, TTA efficiently provides an uncertainty estimate, without the need to train multiple models. This highlights an advantage of the TTA method and should be considered for future applications in surrogate modeling.

\section{Conclusion} 
\label{Conclusion}

The proposed TTA method enhances the prediction accuracy and reliability of RNNs for path-dependent deformations in SFRCs. By averaging multiple augmented predictions, it reduces the MeRE and provides a more consistent time signal shape prediction compared to single predictions. Additionally, the TTA method offers a possible measure of prediction uncertainties, correlating well with the actual prediction errors. Thus, the proposed TTA method provides a valuable tool for confidence estimation in model outputs, further strengthening the reliability of the predictions.

However, the method is not without its potential drawbacks. The proposed error estimation is accurate on average across a dataset, however, a correlation does not ensure that the it corresponds to the uncertainty of the network predictions. For example, for other RNN architectures and types of data, the correlation might not remain. Therefore, the analysis performed does not yet compare with previous methods of uncertainty estimates and a would first have to be verified that the variation of predictions relates to the uncertainty of the RNN.

Moreover, by evaluating the RNN used in this study, it does not yet enforce physical constraints. Therefore, certain predictions could result in unrealistic, or even break the laws of physics. Due to the nature of the RNN, it would struggle with the long-term memory, and such unrealistic predictions could appear more frequently. These considerations highlight areas for potential improvement and optimization.
  
In light of these considerations, future research could focus on expanding the TTA method to other types of neural network architectures, enforcing physics constraints, improving long-term memory, and optimizing the neural network architecture. Such developments could broaden the scope and impact of the method, extend it to other fields, and provide new strategies for the explainability of neural networks in general.


\section*{Acknowledgment}
The authors gratefully acknowledge financial support from the Swedish Research Council (VR grant: 2019-04715) and the University of Gothenburg. 


\bibliographystyle{unsrt}
\bibliography{2.References}{}

\end{document}